\documentclass[prd,aps,tightenlines,a4paper,12pt]{revtex4}

\usepackage{graphicx}
\usepackage{bm}

\usepackage{url}

\newcommand{\OO}[1]{{\mathcal O}(c^{-#1})}

\newcommand{\muas}[0]{\hbox{\rm $\mu$as}}

\newcommand{\ve}[1]{\mbox{\boldmath$#1$}}
\def\source{{\rm 0}}
\def\obs{{\rm 1}}

\begin{document}

\title{Efficient computation of the quadrupole light deflection}

\author{Sven \surname{Zschocke}, Sergei A. \surname{Klioner}}
\affiliation{
Lohrmann Observatory, Dresden Technical University,\\
Mommsen Str. 13, D-01062 Dresden, Germany\\
}

\begin{abstract}
\begin{center}
{\it GAIA-CA-TN-LO-SZ-001-2}

\medskip

\today

\end{center}

Efficient computation of the quadrupole light deflection for both
stars/quasars and solar system objects within the framework of the
baseline Gaia relativity model (GREM) is discussed. 
Two refinements have been achieved with the goal to improve the performance of the model:

\bigskip

{
\leftskip 1.5cm
\parindent -0.4cm

-- The quadrupole deflection formulas for both cases are simplified
as much as possible considering the Gaia nominal orbit (only
approximate minimal distances between Gaia and the giant planets were
used here), physical parameters of the giant planets and the envisaged
accuracy of $1\,\mu{\rm as}$ for individual systematic effects. The
recommended formulas are given by Eq.~(\ref{light_185}) for
stars/quasars and by Eq.~(\ref{source_180}) for solar system objects.

\bigskip

-- Simple expressions for the upper estimate of the quadrupole light
deflection have been found allowing, with a few additional arithmetical
operations, to judge a priori if the quadrupole light deflection should
be computed or not for a given source and for a given requested
accuracy. The recommended criteria are given by Eq.
(\ref{criterion_30}) for stars/quasars and by Eq.
(\ref{criterion_source_60}) for solar system objects.

\bigskip

}

\bigskip

The quadrupole Shapiro effect for solar system objects is reconsidered. 
A strict upper bound of quadrupole Shapiro effect for solar system objects 
is given in Eq.~(\ref{shapiro_150}).

\end{abstract}

\maketitle

\newpage

\tableofcontents

\newpage

\section{Introduction}

Gaia mission will reach an accuracy on microarcsecond ($\mu{\rm as}$) level.
This level of accuracy requires a precise modelling of light propagation. In particular, 
the light deflection due to quadrupole gravitational field of deflecting bodies
should be taken into account \cite{Klioner2}. Analytical formulas for quadrupole light deflection are well 
known. Analytical solutions of light deflection in a quadrupole gravitational field have been investigated by
many authors \cite{Ivanitskaya1979,Epstein_Shapiro1,Richter_Matzner1,Richter_Matzner2,Richter_Matzner3,Cowling1,
Klioner1,Klioner_Kopejkin,Klioner2,Klioner:Blankenburg:2003}. For the first time the full analytical
solution for the light trajectory in a quadrupole field has been obtained in \cite{Klioner1}.
These results were confirmed by a different approach in \cite{LePoncinLafitteTeyssandier2008}. Various
generalization (higher-order multipole moments, time-dependence, etc.)
were derived in \cite{Kopeikin1997,KopeikinKorobkovPolnarev2006,KopeikinMakarov2007}.
The formulas suitable for high-accuracy data reduction are given e.g. in \cite{Klioner2}.

However, the Gaia mission will determine the astrometric positions of about $10^9$ objects, 
implying a data reduction of about $10^{12}$ individual observations during the mission time
of $5$ years. It is, therefore, obvious that efficient analytical solutions of quadrupole light deflection are
mandatory in order to succeed with data reduction.
But the full expressions of these solutions of light deflection are rather involved and much too time-consuming 
for practical Gaia data reduction. Furthermore, the quadrupole light deflection will reach the microarcsecond 
level only for objects within a small observational field of giant planets. Accordingly, it is highly useful 
to find analytical criteria by means of which one can decide whether or not the quadrupole field needs to be taken 
into account. Such criteria can only be obtained by simpler analytical formulas. And finally, the implementation 
of the full expressions involve round-off errors in the Gaia data reduction. Therefore, it is tempting to obtain 
simpler analytical expressions which will not be hampered by such problems. By means of the assumption the Gaia 
spacecraft is located near the Earth's orbit (Gaia will have a Lissajous-like orbit around Lagrange point $L2$), 
we have obtained simpler expressions valid on microarcsecond level of accuracy. Criteria by means of which one 
can decide whether or not it is necessary to take into account the quadrupole effect have been derived. 

Furthermore, the accuracy of radio and laser radar links of future missions like BepiColumbo or Juno require 
modelling of the light travel time at the level of millimeters. Hence, also the Shapiro delay due to quadrupole 
fields is of practical interest. Therefore and for reasons of completeness, the known analytical expressions of 
quadrupole Shapiro effect will be reconsidered. Especially, an improved estimate of this effect is given. 

The report is organized as follows: In Section \ref{Basics} we summarize some basics about light deflection 
and introduce the notation. In Section \ref{StarsandQuasars} the full quadrupole formula in post-Newtonian order 
for stars and quasars is presented. A simplified expression and criteria for stars and quasars 
for quadrupole light deflection are given in Sections \ref{ApproximationStars} and \ref{CriteriaStars}. 
In Section \ref{SolarSystemObjects} the full quadrupole formula in post-Newtonian order for solar system objects 
is presented. A simplified expression and criteria for quadrupole light deflection of solar system objects are 
given in Sections \ref{ApproximationSolar} and \ref{CriteriaSolar}. 
A strict upper bound of quadrupole Shapiro effect is given in Section \ref{Shapiro}. 
Numerical tests are given in Section \ref{NumericalTests}. The findings are summarized in Section \ref{Summary}.

\section{Some basics about light propagation\label{Basics}}

Let us summarize some basic formulas of light propagation in post-Newtonian approximation. The geodetic equations 
in post-Newtonian order is linear with respect to the metric components and, therefore, the coordinates of a 
photon and the derivative with respect to coordinate time $t$ is given by \cite{Klioner2} 
\begin{eqnarray}
\ve{x} (t) &=& \ve{x} (t_{\source}) + c\,\ve{\sigma} \,(t - t_{\source}) + \sum\limits_i \Delta \ve{x}_i (t)\,,
\label{light_path_5}
\\
\nonumber\\
\dot{\ve{x}} (t) &=& c\,\ve{\sigma} + \sum\limits_i\Delta \dot{\ve{x}}_i (t)\,.
\label{light_path_6}
\end{eqnarray}

\noindent
Here, $t_0$ is the time moment of emission, $\ve{x}_{\source} = \ve{x} (t_{\source})$ is the position of the photon
at the moment of emission, i.e. the position of source, and $\ve{\sigma} = \lim_{t\rightarrow - \infty}
\frac{\displaystyle \dot{\ve{x}} (t)}{\displaystyle c}$ is the unit tangent vector at infinitly past.
The position of observer is $\ve{x}_{\obs} = \ve{x} (t_{\obs})$ and $t_{\obs}$ is the moment of observation. 
The unit coordinate direction of the light propagation at the moment of observation reads
$\ve{n} = \frac{\displaystyle \dot{\ve{x}} (t_{\obs})}{\displaystyle \left|\dot{\ve{x}} (t_{\obs})\right|}$.
In post-Newtonian order the transformation $\ve{\sigma}$ to $\ve{n}$ reads
\begin{eqnarray}
\ve{n} &=& \ve{\sigma} + \sum \limits_i \delta \ve{\sigma}_i + {\cal O} \left(c^{-4}\right)\,,
\label{lightpath_A}
\end{eqnarray}

\noindent
where
\begin{eqnarray}
\delta\ve{\sigma}_i=\ve{\sigma}\times
\left( c^{-1}\,\Delta\dot{\ve{x}}_i\left(t_{\obs}\right)\times\ve{\sigma}\right). 
\label{lightpath_C}
\end{eqnarray}

\noindent
The sum in (\ref{lightpath_A}) runs over individual terms in the metric of various physical origins 
(e.g. monopole gravitational field of various bodies, quadrupole fields, higher multipole fields, etc.).
The spherical symmetric term of light deflection is given by 
\begin{eqnarray}
\delta \ve{\sigma}_{\rm pN} (t_{\obs}) &=& \ve{\sigma} \times \left(
c^{-1}\,\Delta_{\rm pN} \, \dot{\ve{x}} (t_{\obs}) \times \ve{\sigma} \right) 
= \sum\limits_A \delta \ve{\sigma}_{\rm pN}^A (t_{\obs}) \,,
\nonumber\\
\nonumber\\
\delta \ve{\sigma}_{\rm pN}^A (t_{\obs}) &=& - \frac{\left(1+\gamma\right)\,G\,M_A}{c^2}\,\frac{\ve{d}_A}{d_A^2} 
\left(1 + \frac{\ve{\sigma}\cdot\ve{r}_{\obs}^A}{r_{\obs}^A}\right).
\label{quadrupole_pN}
\end{eqnarray}

\noindent
The sum in (\ref{quadrupole_pN}) runs over the bodies $A$ of solar system. The impact vector
\begin{eqnarray}
\ve{d}_A &=& \ve{\sigma} \times \left(\ve{r}_{\obs}^A \times \ve{\sigma}\right),
\label{impact_sigma}
\end{eqnarray}

\noindent
has been introduced having the absolute value $d_A = \left|\ve{d}_A\right|$. The vector 
$\ve{r}_{\obs}^A = \ve{x}_{\obs} - \ve{x}_A$ is directed from body $A$ towards the observer; the absolute value 
$r_{\obs}^A = |\ve{r}_{\obs}^A|$, and $\gamma$ is the PPN parameter (for general theory of relativity $\gamma=1$).
We also note the absolute value of the monopole contribution from one body $A$:
\begin{eqnarray}
\left|\,\delta \ve{\sigma}_{\rm pN}^A (t_{\obs}) \,\right| 
&=& \left(1+\gamma\right) \frac{G\,M_A}{c^2}\,\frac{1}{d_A}
\left(1 + \frac{\ve{\sigma}\cdot\ve{r}_{\obs}^A}{r_{\obs}^A}\right) \le 
2\,\left(1+\gamma\right) \frac{G\,M_A}{c^2}\,\frac{1}{d_A}\,.
\label{absolute_quadrupole_pN}
\end{eqnarray}

\noindent
In order to consider light propagation between two given points
$\ve{x}_0$ and $\ve{x}_1$ (as it is needed for the data processing for
solar system objects) let us define the vector $\ve{r}_{\source}^A = \ve{x}_{\source} - \ve{x}_A$, 
which is directed from body $A$ towards the source and the vector
$\ve{R} = \ve{x}_{\obs} - \ve{x}_{\source} = \ve{r}_{\obs}^A - \ve{r}_{\source}^A$ is directed from
source towards observer. Furthermore, the unit direction from source to observer is
$\ve{k} = \frac{\displaystyle \ve{R}}{\displaystyle R}$,
and the absolute value $R = |\ve{R}|$, $r_{\source}^A = |\ve{r}_{\source}^A|$. 

In post-Newtonian order, the transformation $\ve{k}$ to $\ve{n}$ reads
\begin{eqnarray}
\ve{n} &=& \ve{k} + \sum\limits_i \delta \ve{k}_i + {\cal O} \left(c^{-4}\right) \,,
\label{criterion_source_43}
\end{eqnarray}

\noindent
where 
\begin{eqnarray}
\delta\ve{k}_i = \ve{k}\times\left(c^{-1}\,\Delta\dot{\ve{x}}_i\left(t_{\obs} \right) 
- R^{-1}\,\Delta \ve{x}_i \left(t_{\obs}\right)\right)\times\ve{k}\,.
\label{spherical_symmetric_part_k}
\end{eqnarray}

\noindent
The sum in (\ref{criterion_source_43}) runs over individual terms in the metric of various physical origins
(e.g. monopole gravitational field of various bodies, quadrupole fields, higher multipole fields, etc.).
Thus, the spherical symmetric part reads (cf. Eq.~(70) in \cite{Klioner2} or cf. Eq.~(24) in 
\cite{Article_Klioner_Zschocke}): 
\begin{eqnarray}
\delta \ve{k}_{\rm pN} (t_{\obs}) &=& 
\ve{k}\times\left(\left(c^{-1}\,\Delta\dot{\ve{x}}_{\rm pN}\left(t_{\obs}\right) 
- R^{-1}\,\Delta \ve{x}_{\rm pN} \left(t_{\obs}\right) \right)\times\ve{k}\right)
= \sum\limits_A \delta \ve{k}_{\rm pN}^A (t_{\obs}) \,,
\nonumber\\
\nonumber\\
\delta \ve{k}_{\rm pN}^A (t_{\obs}) &=& - \left(1+\gamma\right)\,\frac{G}{c^2}\,
\frac{M_A}{r_{\obs}^A}\,\frac{\ve{k} \times \left(\ve{r}_{\source}^A \times \ve{r}_{\obs}^A\,\right)}
{r_{\source}^A\,r_{\obs}^A + \ve{r}_{\source}^A \cdot \ve{r}_{\obs}^A}\,.
\label{criterion_source_45}
\end{eqnarray}

\noindent
The sum in (\ref{criterion_source_45}) runs over the bodies $A$ of solar system.
We also note the absolute value of monopole term from one body $A$:
\begin{eqnarray}
\left|\,\delta \ve{k}_{\rm pN}^A (t_{\obs}) \right| &=& \left(1+\gamma\right)\,\frac{G}{c^2}\,
\frac{M_A}{r_{\obs}^A}\,\frac{\left|\,\ve{r}_{\source}^A \times \ve{r}_{\obs}^A\,\right|}
{r_{\source}^A\,r_{\obs}^A + \ve{r}_{\source}^A \cdot \ve{r}_{\obs}^A}\,.
\label{criterion_source_50}
\end{eqnarray}

\noindent
Here, the impact parameter $\ve{d}_A$ can be computed as 
\begin{eqnarray}
\ve{d}_A = \ve{k}
\times \left(\ve{r}_{\obs}^A \times \ve{k}\right) +\OO2 = \ve{k}
\times \left(\ve{r}_{\source}^A \times \ve{k}\right)+\OO2\,,
\label{impact_k}
\end{eqnarray}

\noindent
because of $\ve{\sigma} = \ve{k} + {\cal O} \left(c^{-2}\right)$.


\section{The quadrupole light deflection for stars and quasars\label{StarsandQuasars}}

Using the expression $\Delta\dot{\ve{x}}_{\rm Q}\left(t_{\obs}\right)$ given by Eq.~(44) of \cite{Klioner2} 
and inserting into Eq.~(\ref{lightpath_C}) one gets \cite{Klioner:Blankenburg:2003}
\begin{eqnarray}
\delta \ve{\sigma}_{\rm Q} (t_{\obs}) &=& \sum\limits_A \delta \ve{\sigma}_{\rm Q}^A (t_{\obs})\,,  
\nonumber\\
\nonumber\\
\delta \ve{\sigma}_{\rm Q}^A (t_{\obs}) &=& {1+\gamma\over 2} \,\frac{G}{c^2}\,
\left[ \ve{\alpha}_A^{\prime}\,
\frac{\dot{\cal U}_A (t_{\obs})}{c} + \ve{\beta}_A^{\prime}\,
\frac{\dot{\cal E}_A (t_{\obs})}{c} + \ve{\gamma}_A^{\prime}\,
\frac{\dot{\cal F}_A (t_{\obs})}{c} + \ve{\delta}_A^{\prime}\,
\frac{\dot{\cal V}_A (t_{\obs})}{c}\right] \,,
\label{light_35}
\end{eqnarray}

\noindent
where the sum in (\ref{light_35}) runs over the massive bodies $A$ of solar system.
The scalar functions are 
\begin{eqnarray}
\frac{\dot{\cal U}_A}{c} &=& 
\frac{1}{d_A^3} \, \left( \,2 + 3 \, \frac{\ve{\sigma} \cdot \ve{r}_{\obs}^A}{r_{\obs}^A}
- \frac{\left(\ve{\sigma} \cdot \ve{r}_{\obs}^A\right)^3}{\left(r_{\obs}^A\right)^3}\,\right) \,,
\label{light_40}
\\
\frac{\dot{\cal E}_A}{c} &=&
\frac{\left(r_{\obs}^A\right)^2 - 3 \, \left(\ve{\sigma}\cdot \ve{r}_{\obs}^A \right)^2}{\left(r_{\obs}^A\right)^5}\,,
\label{light_45}
\\
\frac{\dot{\cal F}_A}{c} &=& - 3 \, d_A \,
\frac{\ve{\sigma} \cdot \ve{r}_{\obs}^A}{\left(r_{\obs}^A\right)^5}\,,
\label{light_50}
\\
\frac{\dot{\cal V}_A}{c} &=& - \frac{1}{\left(r_{\obs}^A\right)^3}\,,
\label{light_55}
\end{eqnarray}

\noindent
and the time-independent vectorial coefficients
\begin{eqnarray}
\alpha_A^{'\; k} &=& - M_{i j}^A \, \sigma^{i} \, \sigma^j \,
\frac{d_A^k}{d_A} \, + \, 2 \,
M^A_{k j} \, \frac{d_A^{j}}{d_A}
\, - \,  2 \, M^A_{i j} \, \sigma^i \, \sigma^k \, \frac{d_A^j}{d_A}
\, - \, 4 \, M^A_{i j} \, \frac{d_A^i \, d_A^j \, d_A^k}{d_A^3} \,,
\label{light_60}
\\
\beta_{A}^{'\;k} &=& 2 M^A_{i j}\,\sigma^i\,\frac{d_A^j \, d_A^k}{d_A^2} \,,
\label{light_65}
\\
\gamma_A^{'\;k} &=& M^A_{i j} \, \frac{d_A^i \, d_A^j \, d_A^k}{d_A^3} \, - \,
M^A_{i j} \, \sigma^i \, \sigma^j \, \frac{d_A^k}{d_A} \,,
\label{light_70}
\\
\delta_A^{'\;k} &=& - 2 \, M^A_{i j} \, \sigma^i \, \sigma^j \, \sigma^k
\, + \, 2 \, M^A_{k j} \, \sigma^j \,
- \, 4 \, M^A_{i j} \, \sigma^i \, \frac{d_A^j \, d_A^k}{d_A^2} \,.
\label{light_75}
\end{eqnarray}

\noindent
The quadrupole formula (\ref{light_35}) is valid for sources at infinite distance from the observer. The sum over 
$A$ in (\ref{light_35}) runs, in principle, over all bodies inside the solar system, but only the giant planets 
contribute within the accuracy of $1\,\mu{\rm as}$; the quadrupole effect of the Sun is irrelevant for Gaia mission 
because of the $45$ degrees observation angle of the Sun by Gaia. 

The symmetric and tracefree quadrupole moment of an object $A$ is defined in 
\cite{Klioner_Kopejkin,Klioner1,Klioner2,Klioner:Blankenburg:2003} and given by 
\begin{eqnarray}
M^A_{ij} &=&
\int \limits_A d^3 x \, \rho_A (x) \,
\left( r^i \, r^j  - \frac{1}{3} \, \delta^{ij} \, r^2 \right),
\quad r^i = x^i - x_A^i \,,
\label{light_90}
\end{eqnarray}

\noindent
with the mass density $\rho_A$, and the integral is taken over the volume of body $A$; the Kronecker symbol 
$\delta^{i j} =1$ for $i=j$ and zero otherwise. For an axial symmetric body (this approximation is sufficient 
for the giant planets and aimed accuracy of $1 \mu{\rm as}$) one has (see Eqs.~(48) -- (53) of \cite{Klioner2})
\begin{eqnarray}
\begin{array}[t]{l}
\displaystyle
M^A_{i j} =
M_A \, J_2^A \, P_A^2 \, \frac{1}{3} \, {\cal R} \, \left ( \begin{array}[c]{l}
\displaystyle
1 \quad \quad 0 \quad \quad 0 \\
\\
\displaystyle
0 \quad \quad 1 \quad \quad 0 \\
\\
\displaystyle
0 \quad \quad 0 \quad - 2
\end{array}
\right) \,{\cal R}^{T}
\end{array}\,,
\label{light_95}
\end{eqnarray}

\noindent
where ${\cal R}$ is the rotational matrix giving the orientation of the symmetry (rotational) axis $\ve{e}_3$ of the 
massive body in the BCRS, $M_A$ is the mass of the massive body $A$, $J_2^A$ is the coefficient of the second zonal 
harmonic of the gravitational field, $P_A$ is the minimal radius of a sphere containing the body $A$ and whose center 
coincides with the center of mass of $A$ (for the giant planets $P_A$ is just the equatorial radius).


\section{Approximation of the quadrupole light deflection formula for stars and quasars\label{ApproximationStars}}

In this Section we will derive an approximation of (\ref{light_35}) sufficient for the envisaged accuracy of 
$1\,\mu{\rm as}$. From (\ref{light_35}) one obtains the estimate
\begin{eqnarray}
|\delta \ve{\sigma}_{\rm Q} | &\le&
{1+\gamma\over 2}\,\frac{G}{c^2}\,\sum\limits_{A}
\left[\left|\ve{\alpha}_A^{\prime}\right|
\frac{|\dot{\cal U}_A|}{c} + \left| \ve{\beta}_A^{\prime}\right|
\frac{|\dot{\cal E}_A|}{c} + \left| \ve{\gamma}_A^{\prime}\right|
\frac{|\dot{\cal F}_A|}{c} + \left| \ve{\delta}_A^{\prime}\right|
\frac{|\dot{\cal V}_A|}{c} \right] \,.
\label{light_125}
\end{eqnarray}

\noindent
Note that since $\delta \ve{\sigma}_{\rm Q}$ is perpendicular to $\ve{\sigma}$, the absolute value 
$|\,\delta \ve{\sigma}_{\rm Q}\,|$ gives, in the adopted post-Newtonian approximation, the change of the 
calculated or observed direction to star or quasar due to the quadrupole light deflection. Now we will estimate the 
terms in (\ref{light_125}). For such an estimation we consider the case of an axial symmetric body because this 
approximation is sufficient for the giant planets and goal accuracy of $1\,\mu{\rm as}$, i.e. we take the quadrupole 
tensor in the form (\ref{light_95}).

\subsection{Estimate of the vectorial coefficients}

In this Section we estimate the vectorial coefficients of the last three individual terms in (\ref{light_125}). 
For estimating the maximal possible absolute value of the coefficients 
$\left| \ve{\beta}_A^{\prime}\right|$, $\left| \ve{\gamma}_A^{\prime}\right|$ and 
$\left| \ve{\delta}_A^{\prime}\right|$, we replace ${\cal R}$ by the unit matrix.
Then, by inserting (\ref{light_95}) into (\ref{light_60}) - (\ref{light_75}) we obtain 
\begin{eqnarray}
\ve{\alpha}_A^{\prime} &=& - M_A J_2^A P_A^2 \frac{1}{d_A}
\Bigg[ \left(1 - \left(\ve{\sigma} \cdot \ve{e}_3 \right)^2
 - 4 \frac{\left(\ve{d}_A \cdot \ve{e}_3\right)^2}{d_A^2} \right) \ve{d}_A
+ 2 \left(\ve{d}_A \cdot \ve{e}_3 \right) \ve{e}_3 
\nonumber\\
&& - 2 \, \left(\ve{\sigma} \cdot \ve{e}_3 \right)\left(\ve{d}_A \cdot \ve{e}_3\right) \ve{\sigma} \Bigg] \,,
\label{light_105}
\\
\nonumber\\
\ve{\beta}_A^{\prime} &=& - 2\,M_A\,J_2^A\,P_A^2\,\frac{1}{d_A^2}
\left(\ve{\sigma} \cdot \ve{e}_3 \right) \left(\ve{d}_A \cdot \ve{e}_3\right) \ve{d}_A \,,
\label{light_110}
\\
\nonumber\\
\ve{\gamma}_A^{\prime} &=& - M_A \, J_2^A\,P_A^2\,\frac{1}{d_A^3}\,
\left[ \left(\ve{d}_A \cdot \ve{e}_3\right)^2 \ve{d}_A - \left(\ve{\sigma} \cdot \ve{e}_3 \right)^2 d_A^2 \ve{d}_A 
\right] \,,
\label{light_115}
\\
\nonumber\\
\ve{\delta}_A^{\prime} &=& 2 \, M_A \, J_2^A \, P_A^2 \,
\left[ \left(\ve{\sigma} \cdot \ve{e}_3 \right)^2 \ve{\sigma} + \frac{2}{d_A^2} 
\left(\ve{\sigma} \cdot \ve{e}_3 \right) \left(\ve{d}_A \cdot \ve{e}_3\right) \ve{d}_A
- \left(\ve{\sigma} \cdot \ve{e}_3 \right) \ve{e}_3 \right] \, ,
\label{light_120}
\end{eqnarray}

\noindent
where $\ve{e}_3$ is the unit direction along the axis of symmetry (rotation). Here, 
$\left(\ve{\sigma} \cdot \ve{e}_3 \right)$ and $\left(\ve{d}_A \cdot \ve{e}_3\right)$ are the projections of the 
vectors $\ve{\sigma}$ and $\ve{d}_A$, respectively, on the axis of symmetry. With the aid of 
(\ref{light_110}) - (\ref{light_120}) we can explicitly determine the maximal absolute values of the last three 
individual terms in (\ref{light_125}):
\begin{eqnarray}
\left|\ve{\beta}_A^{\prime}\right|
&\le&  2 \, M_A \, \left| J_2^A \right| \, P_A^2 \,
\left| \ve{\sigma} \cdot \ve{e}_3 \right| \,\frac{\ve{d}_A \cdot \ve{e}_3}{d_A}
\le M_A \, \left| J_2^A \right| \, P_A^2 \,,
\label{light_130}
\\
\left|\ve{\gamma}_A^{\prime}\right|
&=&  M_A \, \left| J_2^A \right| \, P_A^2 \,
\left| \frac{\left(\ve{d}_A \cdot \ve{e}_3\right)^2}{d_A^2} - \left( \ve{\sigma} \cdot \ve{e}_3 \right)^2\right|
\le \, M_A \, \left| J_2^A\right| \, P_A^2 \,,
\label{light_135}
\\
\left|\ve{\delta}_A^{\prime}\right|
&\le& 2 \,M_A \left| J_2^A \right| P_A^2 
\left( \left( \ve{\sigma} \cdot \ve{e}_3 \right)^2 + 2 \frac{\left|\left( \ve{\sigma} \cdot \ve{e}_3 \right)
\left(\ve{d}_A \cdot \ve{e}_3\right)\right|}{d_A} 
+ \left( \ve{\sigma} \cdot \ve{e}_3 \right)\right)
\le 6 \,M_A \left| J_2^A \right| P_A^2 \,,
\label{light_140}
\end{eqnarray}

\noindent
where for the estimates (\ref{light_130}) and (\ref{light_140}) we have taken into account that
\begin{eqnarray}
\left|\ve{\sigma} \cdot \ve{e}_3\right|\,\frac{\left|\ve{d}_A \cdot \ve{e}_3\right|}{d_A} &\le& \frac{1}{2} \,,
\label{light_142}
\end{eqnarray}

\noindent
valid due to $ \ve{\sigma} \cdot \ve{d}_A = 0$.

\subsection{Estimate of the scalar functions}

Furthermore, from (\ref{light_45}) - (\ref{light_55}) we deduce the estimates
\begin{eqnarray}
\frac{|\dot{\cal E}_A|}{c} &\le& \frac{\left(r_{\obs}^A\right)^2
+ 3\,\left(r_{\obs}^A \right)^2}{\left(r_{\obs}^A\right)^5} = 4 \, \frac{1}{\left(r_{\obs}^A\right)^3}\,,
\label{light_145}
\\
\nonumber\\
\frac{|\dot{\cal F}_A|}{c} &\le& d_A\,
\frac{3\,r_{\obs}}{\left(r_{\obs}^A\right)^5} \le 3\,\frac{1}{\left(r_{\obs}^A\right)^3}\,,
\label{light_150}
\\
\nonumber\\
\frac{|\dot{\cal V}_A|}{c} &=& \frac{1}{\left(r_{\obs}^A\right)^3}\,.
\label{light_155}
\end{eqnarray}

\noindent
By inserting these estimates (\ref{light_130}) - (\ref{light_140}) and (\ref{light_145}) - (\ref{light_155}) 
into (\ref{light_125}) we get
\begin{eqnarray}
\frac{G}{c^2} \,
\left|\ve{\beta}_A^{\prime}\right| \, \frac{|\dot{\cal E}_A|}{c}
&\le& 4 \, \frac{G}{c^2}\,M_A\,\left|J_2^A\right| \,P_A^2\,\frac{1}{\left(r_{\obs}^{A\;{\rm min}}\right)^3}\,,
\label{light_160}
\end{eqnarray}

\begin{eqnarray}
\frac{G}{c^2} \,
\left|\ve{\gamma}_A^{\prime}\right| \, \frac{|\dot{\cal F}_A|}{c}
&\le& 3 \, \frac{G}{c^2}\,M_A\,\left|J_2^A\right|\,P_A^2\,\frac{1}{\left(r_{\obs}^{A\;{\rm min}}\right)^3} \,,
\label{light_165}
\end{eqnarray}

\begin{eqnarray}
\frac{G}{c^2} \,
\left|\ve{\delta}_A^{\prime}\right| \, \frac{|\dot{\cal V}_A|}{c}
&\le& 6 \, \frac{G}{c^2}\,M_A\,\left|J_2^A\right|\,P_A^2\,\frac{1}{\left(r_{\obs}^{A\;{\rm min}}\right)^3}\,.
\label{light_170}
\end{eqnarray}

\noindent
The quantity $r_{\obs}^{A\;{\rm min}}$ represents the minimal distance between the object $A$ and the observer.

\subsection{Collection of all terms}

\begin{table}[t!]
\begin{tabular}{c | c | c | c | c}
&&&&\\[-10pt]
Parameter &\hbox to 2mm{} Jupiter \hbox to 2mm{} & \hbox to 2mm{} Saturn \hbox to 2mm{}& \hbox to 2mm{} Uranus \hbox to 2mm{} & \hbox to 2mm{} Neptune \hbox to 2mm{} \\[3pt]
\hline
&&&&\\[-10pt]
$GM_A/c^2$\  [m] & $1.40987$  & $0.42215$ & $0.064473$ & $0.076067$ \\[3pt]
$J_2^A\  [10^{-3}]$    & $14.697$   & $16.331$ & $3.516$ & $3.538$ \\[3pt]
$P_A$\  [$10^6$ m]  & $71.492$ & $60.268$ & $25.559$ & $24.764$ \\[3pt]
$r_{o A}^{\rm min}$\  [$10^{12}$ m] & $0.59$   & $1.20$ & $2.59$ & $4.31$ \\[3pt]
\hline
&&&&\\[-10pt]
$GM_A\,J_2^A\,P_A^2/c^2$\ [$10^{15}$ m$^3$] & $0.106$ & $0.025$
& $0.000148$ & $0.000165$ \\[3pt]
\end{tabular}
\caption{Numerical parameters of the giant planets taken from \cite{Encyclopedia,IERS2003}.}
\label{table1}
\end{table}

Table~\ref{table1} summarizes physical parameters of the giant planets. In this Table and in the following discussions 
we use values of $r_{\obs}^{A\;{\rm min}}$ computed under assumption that the observer is within a few million 
kilometers from the Earth's orbit. From the values given in Table~\ref{table1} and 
(\ref{light_160}) - (\ref{light_170}) we deduce (for these estimates $\gamma=1$)
\begin{eqnarray}
\frac{G}{c^2}  \, \left[
\left|\ve{\beta}_A^{\prime}\right| \,\frac{|\dot{\cal E}_A|}{c}
+ \left|\ve{\gamma}_A^{\prime}\right| \,\frac{|\dot{\cal F}_A|}{c}
+ \left|\ve{\delta}_A^{\prime}\right| \,\frac{|\dot{\cal V}_A|}{c}
\right]
&\le& 1.61 \times 10^{-9} \, \mu{\rm as} \quad {\rm for} \, {\rm Jupiter}\,,
\nonumber\\
&\le& 4.52 \times 10^{-11} \, \mu{\rm as} \quad {\rm for} \, {\rm Saturn}\,,
\nonumber\\
&\le& 2.64 \times 10^{-14} \, \mu{\rm as} \quad {\rm for} \, {\rm Uranus}\,,
\nonumber\\
&\le& 7.66 \times 10^{-15} \, \mu{\rm as} \quad {\rm for} \, {\rm Neptune}\,.
\label{light_180}
\end{eqnarray}

\noindent
Obviously, by comparing the estimates given in (\ref{light_180}) with the envisaged accuracy of $1 \mu{\rm as}$ we can 
conclude that these last three terms in (\ref{light_35}), i.e. $\left|\ve{\beta}_A^{\prime}\right| \,\dot{\cal E}_A$ 
and $\left|\ve{\gamma}_A^{\prime}\right| \,\dot{\cal F}_A$ and $\left|\ve{\delta}_A^{\prime}\right| \, \dot{\cal V}_A$, 
can safely be neglected. Accordingly, for Gaia mission, the simplified quadrupole light deflection for stars and 
quasars valid on microarcsecond level of accuracy, reads 
\begin{eqnarray}
\delta \ve{\sigma}_{\rm Q} &=& {1+\gamma\over 2}\,\frac{G}{c^2}\,\sum\limits_{A}\,\ve{\alpha}_A^{\prime}\,
\frac{\dot{\cal U}_A}{c} \,,
\label{light_185}
\end{eqnarray}

\noindent
with $\dot{\cal U}_A$ given by (\ref{light_40}) and $\ve{\alpha}_A^{\prime}$ given by (\ref{light_60}).


\section{An upper estimate of the quadrupole light deflection for stars and quasars\label{CriteriaStars}}

The simplified expression of quadrupole light deflection (\ref{light_185}) it still complicated. 
In order to avoid evaluation of this term for each object in the data reduction, a simple 
criterion is needed which allows one, with a few additional arithmetical operations, to judge if the quadrupole 
light deflection should be computed for a given source and for a given accuracy. To deduce such a criterion, 
we first evaluate the absolute value of the vectorial coefficient (\ref{light_105}),
\begin{eqnarray}
\left|\ve{\alpha}_A^{\prime}\right| &=& M_A \, \left|J_2^A\right| \, P_A^2
\left(1 - \left(\ve{\sigma} \cdot \ve{e}_3\right)^2 \right) \,,
\label{criterion_5}
\end{eqnarray}

\noindent
which yields for the absolute value of light deflection angle caused by the quadrupole field of an massive 
object $A$ 
\begin{eqnarray}
| \delta \ve{\sigma}_{\rm Q}^A | &=& {1+\gamma\over 2} \,
\frac{GM_A}{c^2} \,\left|J_2^A\right| \, P_A^2 
\left(1 - \left(\ve{\sigma} \cdot \ve{e}_3\right)^2\right)
\frac{1}{d_A^3} \, \left(2 + 3\,\frac{\ve{\sigma} \cdot \ve{r}_{\obs}^A}{r_{\obs}^A} 
- \frac{\left(\ve{\sigma} \cdot \ve{r}_{\obs}^A\right)^3}{\left(r_{\obs}^A\right)^3} \right).
\label{criterion_10}
\end{eqnarray}

\noindent
A comparison of (\ref{criterion_10}) with the absolute value of spherically symmetric part given in 
(\ref{absolute_quadrupole_pN}) and taking into account the fact
\noindent
\begin{eqnarray}
 3\, \frac{\ve{\sigma} \cdot \ve{r}_{\obs}^A}{r_{\obs}^A} 
- \frac{\left(\ve{\sigma} \cdot \ve{r}_{\obs}^A\right)^3}{\left(r_{\obs}^A\right)^3} + 2
\le \frac{9}{4} \, \left(1 + \frac{\ve{\sigma} \cdot \ve{r}_{\obs}^A}{r_{\obs}^A}\right),
\label{criterion_20}
\end{eqnarray}

\noindent
we obtain the criterion
\begin{eqnarray}
|\,\delta \ve{\sigma}_{\rm Q}^A| &\le&
\frac{9}{8} \, \frac{P_A^2}{d_A^2} \, \left| J_2^A \right| \,
\left( 1 - \left( \ve{\sigma} \cdot \ve{e}_3 \right)^2 \right)
|\,\delta \ve{\sigma}_{\rm pN}^A\,| \,.
\label{criterion_25}
\end{eqnarray}

\noindent
Due to $1 \ge \left( \ve{\sigma} \cdot \ve{e}_3 \right)^2 $, the estimate (\ref{criterion_25}) can be further 
approximated by
\begin{eqnarray}
\left|\,\delta \ve{\sigma}_{\rm Q}^A\right| &\le& \frac{9}{8} \, \left| J_2^A \right| \,
\frac{P_A^2}{d_A^2} \,
\left|\,\delta \ve{\sigma}_{\rm pN}^A\right| \,.
\label{criterion_30}
\end{eqnarray}

\noindent
This criterion relates the quadrupole light deflection for stars and quasars
to the simpler case of spherically symmetric part given in (\ref{absolute_quadrupole_pN}).
It is recommended for Gaia to use (\ref{criterion_30}) as a criterion if the
quadrupole light deflection has to be computed for a given star or quasar.
Eq.~(\ref{criterion_10}) can be used to estimate $\left|\,\delta \ve{\sigma}_{\rm Q}^A \right|$ directly:
\begin{eqnarray}
\left|\,\delta \ve{\sigma}_{\rm Q}^A\right|
&\le& {2\,(1+\gamma)\,GM_A\over c^2}\,{P_A^2\over d_A^3}\,\left|J_2^A\right|
\le{2\,(1+\gamma)\,GM_A\over c^2\,P_A}\,\left|J_2^A\right|\,,
\label{criterion_32}
\end{eqnarray}

\noindent
where we have used $\left|2 + 3\,\cos \alpha - \cos^3 \alpha \right| \le 4$. 
The estimate (\ref{criterion_32}) coincides with \cite{Klioner1} (see Eq.~(41) and the sentence below 
in that reference).


\section{The quadrupole light deflection for solar system objects\label{SolarSystemObjects}}

The quadrupole light deflection for solar system objects $\delta \ve{k}_{\rm Q}$ is defined by Eqs.~(36)--(47) 
and (69) of \cite{Klioner2}. Using Eq.~(\ref{spherical_symmetric_part_k}) it can be written as 
\cite{Klioner:Blankenburg:2003}:
\begin{eqnarray}
\delta \ve{k}_{\rm Q} &=& \sum\limits_{A} \delta \ve{k}_{\rm Q}^A\,,
\nonumber\\
\nonumber\\
\delta \ve{k}_{\rm Q}^A &=&{1+\gamma\over 2}\,  \frac{G}{c^2}\,
\left[\ve{\alpha}_A^{\prime \prime}
\frac{{\cal A}_A (t_{\obs})}{c} + \ve{\beta}_A^{\prime \prime}\,
\frac{{\cal B}_A (t_{\obs})}{c} + \ve{\gamma}_A^{\prime \prime}\,
\frac{{\cal C}_A (t_{\obs})}{c} + \ve{\delta}_A^{\prime \prime} \,
\frac{{\cal D}_A (t_{\obs})}{c}\right] \,,
\label{source_5}
\end{eqnarray}

\noindent
where the sum in (\ref{source_5}) runs over the massive bodies $A$ of solar system.
The scalar functions are 
\begin{eqnarray}
\frac{{\cal A}_A}{c} &=&  \frac{1}{d_A} \, \frac{1}{R}
\left(\frac{1}{r_{\source}^A} \frac{r_{\source}^A + \ve{k} \cdot \ve{r}_{\source}^A}
{r_{\source}^A - \ve{k} \cdot \ve{r}_{\source}^A}
- \frac{1}{r_{\obs}^A} \frac{r_{\obs}^A + \ve{k} \cdot \ve{r}_{\obs}^A}{r_{\obs}^A - \ve{k} \cdot \ve{r}_{\obs}^A} \right)
+ \frac{d_A}{\left(r_{\obs}^A\right)^3} \,
\frac{2 \, r_{\obs}^A - \ve{k} \cdot \ve{r}_{\obs}^A}{\left(r_{\obs}^A - \ve{k} \cdot \ve{r}_{\obs}^A\right)^2}\,,
\label{source_10}
\\
\frac{{\cal B}_A}{c} &=& \frac{1}{R}
\left(\frac{\ve{k} \cdot \ve{r}_{\source}^A} {\left(r_{\source}^A\right)^3}
- \frac{\ve{k} \cdot \ve{r}_{\obs}^A} {\left(r_{\obs}^A\right)^3} \right)
+ \frac{\left(r_{\obs}^A\right)^2 - 3 \, \left(\ve{k} \cdot \ve{r}_{\obs}^A\right)^2}{\left(r_{\obs}^A\right)^5}\,,
\label{source_15}
\\
\frac{{\cal C}_A}{c} &=& \frac{d_A}{R}
\left(\frac{1}{\left(r_{\source}^A\right)^3} - \frac{1}{\left(r_{\obs}^A\right)^3} \right) - 3 \, d_A \,
\frac{\ve{k} \cdot \ve{r}_{\obs}^A}{\left(r_{\obs}^A\right)^5}\,,
\label{source_20}
\\
\frac{{\cal D}_A}{c} &=&  - \frac{1}{d_A^2} \, \frac{1}{R}
\left(\frac{\ve{k}\cdot\ve{r}_{\source}^A} {r_{\source}^A} - \frac{\ve{k}\cdot\ve{r}_{\obs}^A} {r_{\obs}^A} \right)- \frac{1}{\left(r_{\obs}^A\right)^3}\,,
\label{source_25}
\end{eqnarray}

\noindent
and the time-independent vectorial coefficients are
\begin{eqnarray}
\alpha_A^{\prime\prime\; k} &=& - M_{i j}^A \, k^{i} \, k^j \, \frac{d_A^k}{d_A} \,
\, + \, 2 \,
M^A_{k j} \, \frac{d_A^{j}}{d_A}
\, - \,  2 \, M^A_{i j} \, k^i \, k^k \, \frac{d_A^j}{d_A}
\, - \, 4 \, M^A_{i j} \, \frac{d_A^i \, d_A^j \, d^k}{d_A^3} \,,
\label{source_30}
\\
\nonumber\\
\beta_{A}^{\prime\prime\;k} &=& 2 M^A_{i j} \, k^i \,
\frac{d_A^j \, d_A^k}{d_A^2} \,,
\label{source_35}
\\
\nonumber\\
\gamma_A^{\prime\prime\;k} &=& M^A_{i j} \, \frac{d_A^i \, d_A^j \, d_A^k}{d_A^3}
\, - \, M^A_{i j} \, k^i \, k^j \, \frac{d_A^k}{d_A} \,,
\label{source_40}
\\
\nonumber\\
\delta_A^{\prime\prime\;k} &=& - 2 \, M^A_{i j} \, k^i \, k^j \, k^k
\, + \, 2 \, M^A_{k j} \, k^j \,
- \, 4\, M^A_{i j} \, k^i \, \frac{d_A^j \, d_A^k}{d_A^2} \,.
\label{source_45}
\end{eqnarray}

\noindent
In the following we will investigate how
(\ref{source_5}) can be simplified for a goal accuracy of $1\,\mu{\rm as}$ and taking into account that in 
the case of Gaia the observer is situated within a few million kilometers from the Earth's orbit.

\section{Approximation of quadrupole light deflection for solar system objects\label{ApproximationSolar}}

To determine the magnitude of the individual terms in (\ref{source_5}) we first notice the estimate 
\begin{eqnarray}
|\,\delta \ve{k}_{\rm Q}|
&\le& {1+\gamma\over 2}\, \frac{G}{c^2}\sum\limits_{A} \, 
\left(\left|\ve{\alpha}_A^{\prime \prime}\right|\,\frac{|{\cal A}_A|}{c} 
+ \left|\ve{\beta}_A^{\prime \prime}\right|\,\frac{|{\cal B}_A|}{c} 
+ \left|\ve{\gamma}_A^{\prime \prime}\right|\,\frac{|{\cal C}_A|}{c} 
+ \left|\ve{\delta}_A^{\prime \prime}\right|\,\frac{|{\cal D}_A|}{c}\right).
\label{source_65}
\end{eqnarray}

\noindent
Since $\delta \ve{k}_{\rm Q}$ is perpendicular to $\ve{k}$ the absolute value $|\,\delta \ve{k}_{\rm Q} (t_{\obs})|$ 
gives, in the adopted post-Newtonian approximation, the change of the calculated or observed direction to a 
solar system object due to the quadrupole light deflection. 

\subsection{Estimate of the vectorial coefficients}

In order to estimate the maximal value of vectorial coefficients 
we make use of the diagonalized form of quadrupole moment given in (\ref{light_95}), which yields for the vectorial 
coefficients (\ref{source_30}) - (\ref{source_45})
\begin{eqnarray}
\ve{\alpha}_A^{\prime \prime} &=& - M_A \, J_2^A \, P_A^2 \, \frac{1}{d}_A
\Bigg[ \left(1 - \left(\ve{k} \cdot \ve{e}_3\right)^2
 - 4 \frac{\left(\ve{d}_A \cdot \ve{e}_3\right)^2}{d_A^2} \right) \ve{d}_A
\nonumber\\
\nonumber\\
&& + 2 \left(\ve{d}_A \cdot \ve{e}_3\right)\ve{e}_3
- 2 \left(\ve{k} \cdot \ve{e}_3\right) \left(\ve{d}_A \cdot \ve{e}_3\right) \ve{k}
+ \frac{2}{3} \left(\ve{k} \cdot \ve{d}_A\right)\ve{k} \Bigg],
\label{source_70}
\\
\nonumber\\
\ve{\beta}_A^{\prime \prime} &=& - 2 \, M_A \, J_2^A \, P_A^2 \,
\frac{1}{d_A^2}
\left[ \left(\ve{k} \cdot \ve{e}_3\right)\left(\ve{d}_A \cdot \ve{e}_3\right)\ve{d}_A
- \frac{1}{3} \left(\ve{k} \cdot\ve{d}_A\right)\ve{d}_A \right] \,,
\label{source_75}
\\
\nonumber\\
\ve{\gamma}_A^{\prime \prime} &=& - M_A \, J_2^A \, P_A^2\, \frac{1}{d_A^3}
\left[\left(\ve{d}_A \cdot \ve{e}_3\right)^2 \ve{d}_A - \left(\ve{k}\cdot\ve{e}_3\right)^2 d_A^2\,\ve{d}_A\right]\,,
\label{source_80}
\\
\nonumber\\
\ve{\delta}^{\prime \prime} &=& 2 \, M_A \, J_2^A \, P_A^2 \,
\left[\left(\ve{k} \cdot \ve{e}_3\right)^2 \ve{k} + \frac{2}{d_A^2} \left(\ve{k} \cdot \ve{e}_3\right)
\left(\ve{d}_A \cdot \ve{e}_3\right)\ve{d}_A - \left(\ve{k} \cdot \ve{e}_3\right) \ve{e}_3 
- \frac{2}{3}\,\frac{1}{d_A^2} \left(\ve{k} \cdot\ve{d}_A\right)\ve{d}_A\right] \, ,
\nonumber\\
\label{source_85}
\end{eqnarray}

\noindent
where $\ve{k} \cdot \ve{e}_3$ and $\ve{d}_A \cdot \ve{e}_3$ are the projections
of the vectors $\ve{k}$ and $\ve{d}_A$, respectively, on the axis of symmetry.

From (\ref{source_75}) - (\ref{source_85}) we deduce the following absolute values for the last three vectorial 
coefficients,
\begin{eqnarray}
\left|\ve{\beta}_A^{\prime \prime}\right| 
&\le& 2 \, M_A \, \left| J_2^A \right| \, P_A^2 \,
\left( \left| \ve{k} \cdot \ve{e}_3 \right| \,\frac{\left|\ve{d}_A \cdot \ve{e}_3\right|}{d_A} 
+ \frac{1}{3} \frac{\left|\ve{k} \cdot \ve{d}_A\right|}{d_A}\right)
\, \le \, M_A \, \left| J_2^A \right| \, P_A^2 \,,
\label{source_90}
\\
\nonumber\\
\left|\ve{\gamma}_A^{\prime \prime}\right|
&=& M_A \, \left| J_2^A \right| \, P_A^2 \,
\left| \frac{\left|\ve{d}_A \cdot \ve{e}_3\right|^2}{d_A^2}
- \left|\ve{k} \cdot \ve{e}_3\right|^2\right|
\le \,  M_A \, \left| J_2^A \right| \, P_A^2 \,,
\label{source_95}
\\
\nonumber\\
\left|\ve{\delta}_A^{\prime \prime}\right|
&=& 2 \, M_A \, \left| J_2^A \right| \, P_A^2 \,
\bigg[ \frac{4}{3} \, \frac{1}{d_A^2} \,
\left( 3 \, \left(\ve{k} \cdot \ve{e}_3\right)^3 -
\left(\ve{k} \cdot \ve{e}_3\right) \right) \left(\ve{k} \cdot \ve{d}_A\right) \left(\ve{d}_A \cdot \ve{e}_3\right)
\nonumber\\
&& \hspace{2.0cm}
-\frac{4}{3} \, \frac{1}{d_A^2} \,
\left(\ve{k} \cdot \ve{d}_A\right)^2 \left(\ve{k} \cdot\ve{e}_3\right)^2 
-  \left(\ve{k} \cdot\ve{e}_3\right)^4 + \left(\ve{k} \cdot\ve{e}_3\right)^2 
+ \frac{4}{9} \, \frac{1}{d_A^2} \,\left(\ve{k} \cdot \ve{d}_A\right)^2\bigg]^{1/2}
\nonumber\\
&\le& M_A \, \left| J_2^A \right| \, P_A^2 \,,
\label{source_100}
\end{eqnarray}

\noindent
where for the estimates (\ref{source_90}) and (\ref{source_100}) we have taken into account that
\begin{eqnarray}
\left|\ve{k} \cdot \ve{e}_3 \right|\,\frac{\left|\ve{d}_A \cdot \ve{e}_3 \right|}{d_A} &\le& \frac{1}{2} \,,
\label{source_103}
\\
\nonumber\\
\left(\ve{k} \cdot \ve{e}_3 \right)^2 - \left(\ve{k} \cdot \ve{e}_3 \right)^4 &\le& \frac{1}{4} \,.
\label{source_104}
\end{eqnarray}

\noindent
The first estimate uses the fact that in post-Newtonian order $\ve{k} \cdot \ve{d}_A = 0$.

\subsection{Estimate of the scalar funtions}

In the following we estimate the magnitude of the scalar functions (\ref{source_15}) - (\ref{source_25}).

\subsubsection{Estimate of ${\cal B}_A$}

The coefficient given in (\ref{source_15}) can be written as follows,
\begin{eqnarray}
\frac{{\cal B}_A}{c} &=& \frac{1}{\sqrt{\left(r_{\source}^A\right)^2 + \left(r_{\obs}^A\right)^2 
- 2 \,\ve{r}_{\source}^A \cdot \ve{r}_{\obs}^A}}
\left( \frac{\ve{k} \cdot \ve{r}_{\source}^A}{\left(r_{\source}^A\right)^3} 
- \frac{\ve{k} \cdot \ve{r}_{\obs}^A}{\left(r_{\obs}^A\right)^3} \right) 
+  \frac{\left(r_{\obs}^A\right)^2 - 3 \left(\ve{k} \cdot \ve{r}_{\obs}^A\right)^2}{\left(r_{\obs}^A\right)^5} \,.
\label{source_105}
\end{eqnarray}

\noindent
Inserting the definition of vector $\ve{k}$ yields
\begin{eqnarray}
\frac{{\cal B}_A}{c} &=& \frac{1}{\left(r_{\source}^A\right)^2 + \left(r_{\obs}^A\right)^2 
- 2 \,\ve{r}_{\source}^A \cdot \ve{r}_{\obs}^A}
\left(\frac{r_{\source}^A}{\left(r_{\obs}^A\right)^2} \, \cos \alpha
+ \frac{r_{\obs}^A}{\left(r_{\source}^A\right)^2} \, \cos \alpha - \frac{1}{r_{\source}^A}
- \frac{1}{r_{\obs}^A} \right)
\nonumber\\
&& + \frac{\left(r_{\obs}^A\right)^2 - 3 \left(\ve{k} \cdot \ve{r}_{\obs}^A\right)^2}{\left(r_{\obs}^A\right)^5}\,,
\label{source_110}
\end{eqnarray}

\noindent
where $\cos \alpha = \frac{\displaystyle \ve{r}_{\obs}^A \cdot \ve{r}_{\source}^A}
{\displaystyle r_{\obs}^A\,r_{\source}^A}$. By means of the inequality (with $x=r_{\source}^A, y=r_{\obs}^A$)
\begin{eqnarray}
\left| \frac{1}{x^2 + y^2 - 2 \, x \, y \, \cos \alpha} \,
\left(\frac{x}{y^2} \, \cos \alpha + \frac{y}{x^2} \, 
\cos \alpha - \frac{1}{x} - \frac{1}{y} \right) \right|
\le \frac{x + y}{x^2 \, y^2}
\label{source_115}
\end{eqnarray}

\noindent
valid for any $x\ge0$ and $y\ge0$, we obtain the estimate
\begin{eqnarray}
\frac{|{\cal B}_A|}{c} &\le& \frac{r_{\source}^A + r_{\obs}^A}{\left(r_{\source}^A\right)^2\,\left(r_{\obs}^A\right)^2}
\, + \, \frac{4}{\left(r_{\obs}^A\right)^3} \, \le \, \frac{1}{d_A \, \left(r_{\obs}^A\right)^2} \, + \,
\frac{1}{d_A^2 \, r_{\obs}^A} \, + \, \frac{4}{\left(r_{\obs}^A\right)^3} \,.
\label{source_120}
\end{eqnarray}

\subsubsection{Estimate of ${\cal C}_A$}

The coefficient given in (\ref{source_20}) can be written as follows,
\begin{eqnarray}
\frac{{\cal C}_A}{c} &=& \frac{d_A}{\left(r_{\source}^A\right)^3 \, \left(r_{\obs}^A\right)^3} \,
\frac{\left(r_{\obs}^A\right)^3 - \left(r_{\source}^A\right)^3}
{\sqrt{\left(r_{\source}^A\right)^2 + \left(r_{\obs}^A\right)^2 
- 2 \,\ve{r}_{\source}^A \cdot \ve{r}_{\obs}^A}}
- 3 \, d_A \,\frac{\ve{k} \cdot \ve{r}_{\obs}^A}{\left(r_{\obs}^A\right)^5} \,.
\label{source_125}
\end{eqnarray}

\noindent
Since
\begin{eqnarray}
\frac{1}{\sqrt{\left(r_{\source}^A\right)^2 + \left(r_{\obs}^A\right)^2 - 2 \,\ve{r}_{\source}^A \cdot 
\ve{r}_{\obs}^A}} &\le& \frac{1}{\sqrt{(r_{\source}^A - r_{\obs}^A)^2}}\,,
\label{source_130}
\end{eqnarray}

\noindent
we find for the absolute value
\begin{eqnarray}
\frac{|{\cal C}_A|}{c} &\le& \frac{d_A}{\left(r_{\source}^A\right)^3 \, \left(r_{\obs}^A\right)^3} \,
\frac{| \left(r_{\source}^A\right)^3 - \left(r_{\obs}^A\right)^3 |}{|r_{\source}^A - r_{\obs}^A|}
+ 3 \, \frac{d_A}{\left(r_{\obs}^A\right)^4}\,.
\label{source_135}
\end{eqnarray}

\noindent
By means of the inequality
\begin{eqnarray}
\frac{|x^3 - y^3|}{|x - y|} &\le& \frac{3}{2} \, \left( x^2 + y^2 \right)
\label{source_140}
\end{eqnarray}

\noindent
that is valid for any $x$ and $y$, we obtain the estimate
\begin{eqnarray}
\frac{|{\cal C}_A|}{c} &\le& \frac{3}{2} \, d_A \,
\frac{\left(r_{\source}^A\right)^2+\left(r_{\obs}^A\right)^2}{\left(r_{\source}^A\right)^3\,\left(r_{\obs}^A\right)^3}
 + 3 \, \frac{d_A}{\left(r_{\obs}^A\right)^4} \, \le \,
\frac{3}{2} \, \frac{1}{\left(r_{\obs}^A\right)^3} + \frac{3}{2} \, \frac{1}{d_A^2 \, r_{\obs}^A}
+ 3 \, \frac{d_A}{\left(r_{\obs}^A\right)^4} \,.
\label{source_145}
\end{eqnarray}

\subsubsection{Estimate of ${\cal D}_A$}

The coefficient given in (\ref{source_25}) can be written as follows,
\begin{eqnarray}
\frac{{\cal D}_A}{c} &=& - \frac{1}{d_A^2} \,
\frac{1}{\sqrt{\left(r_{\source}^A\right)^2 + \left(r_{\obs}^A\right)^2 - 2 \,\ve{r}_{\source}^A\cdot\ve{r}_{\obs}^A}}
\left(\frac{\ve{k}\cdot\ve{r}_{\source}^A}{r_{\source}^A} - \frac{\ve{k} \cdot \ve{r}_{\obs}^A}{r_{\obs}^A} \right)
- \frac{1}{\left(r_{\obs}^A\right)^3}\,.
\label{source_150}
\end{eqnarray}

\noindent
Inserting the definition of vector $\ve{k}$ yields
\begin{eqnarray}
\frac{{\cal D}_A}{c} &=& - \frac{1}{d_A^2} \,
\frac{r_{\source}^A\, \cos \alpha + r_{\obs}^A \,\cos \alpha - r_{\source}^A - r_{\obs}^A}
{\left(r_{\source}^A\right)^2 + \left(r_{\obs}^A\right)^2 - 2 \,\ve{r}_{\source}^A \cdot \ve{r}_{\obs}^A}
- \frac{1}{\left(r_{\obs}^A\right)^3} \,.
\label{source_155}
\end{eqnarray}

\noindent
With the aid of the inequality
\begin{eqnarray}
\left| \frac{x\,\cos \alpha + y \, \cos \alpha - x - y}
{x^2 + y^2 - 2 \, x \, y \,\cos \alpha} \right| &\le& \frac{2}{x + y}
\label{source_160}
\end{eqnarray}

\noindent
valid for $x\ge 0$ and $y\ge 0$, we obtain the estimate
\begin{eqnarray}
\frac{|{\cal D}_A|}{c} &\le& 2 \, \frac{1}{d_A^2} \, \frac{1}{r_{\obs}^A} + \frac{1}{\left(r_{\obs}^A\right)^3}\,.
\label{source_165}
\end{eqnarray}

\subsection{Collection of all terms}

Altogether, by inserting the estimates of vectorial coefficients, (\ref{source_90}) - (\ref{source_100}), 
and the scalar coefficients (\ref{source_120}), (\ref{source_145}), (\ref{source_165}) into (\ref{source_65}) yields
\begin{eqnarray}
&& \frac{G}{c^2}
\left(\left|\ve{\beta}_A^{\prime \prime}\right|\,\frac{|{\cal B}_A|}{c} 
+ \left|\ve{\gamma}_A^{\prime \prime}\right|\,\frac{|{\cal C}_A|}{c} 
+ \left|\ve{\delta}_A^{\prime \prime}\right|\,\frac{|{\cal D}_A|}{c}\right)
\nonumber\\
&& \le \, \frac{G}{c^2} \, M_A \, \left| J_2^A \right| \, P_A^2
\,  \left[ \frac{9}{2} \frac{1}{d_A^2\, r_{\obs}^A}
\, + \, \frac{1}{d_A \, \left(r_{\obs}^A\right)^2} \, + \,
\frac{13}{2} \frac{1}{\left(r_{\obs}^A\right)^3} \, + \, 3 \frac{d_A}{\left(r_{\obs}^A\right)^4}
\right]
\nonumber\\
&& \le \, \frac{G}{c^2} \, M_A \, \left| J_2^A \right| \, P_A^2
\,  \left[ \frac{9}{2} \frac{1}{P_A^2\, r_{\obs}^{A\;{\rm min}}}
\, + \, \frac{1}{P_A \, \left(r_{\obs}^{A\;{\rm min}}\right)^2}
\, + \, \frac{19}{2} \frac{1}{\left(r_{\obs}^{A\;{\rm min}}\right)^3} \right] \,,
\label{source_170}
\end{eqnarray}

\noindent
where we have used that $P_A \le d_A \le r_{\obs}^A$. Note, in the last
line of (\ref{source_170}) the first term in the brackets is at least
by a factor of $\sim 10^{4}$ larger than the other two terms. Using
the parameters given in Table \ref{table1} we obtain for the giant planets
($\gamma$ can be safely set to unity for these estimates)
\begin{eqnarray}
\frac{G}{c^2} \left(\left|\ve{\beta}_A^{\prime \prime}\right|\,\frac{|{\cal B}_A|}{c}
+ \left|\ve{\gamma}_A^{\prime \prime}\right|\,\frac{|{\cal C}_A|}{c}
+ \left|\ve{\delta}_A^{\prime \prime}\right|\,\frac{|{\cal D}_A|}{c}\right)
&& \le \, 3.26 \times 10^{-2} \mu{\rm as} \quad {\rm for} \, {\rm Jupiter}\,,
\nonumber\\
&& \le \, 5.32 \times 10^{-3} \mu{\rm as} \quad {\rm for} \, {\rm Saturn}\,,
\nonumber\\
&& \le \, 8.11 \times 10^{-4} \mu{\rm as} \quad {\rm for} \, {\rm Uranus}\,,
\nonumber\\
&& \le \, 5.79 \times 10^{-5} \mu{\rm as} \quad {\rm for} \, {\rm Neptune}\,.
\label{source_175}
\end{eqnarray}

\noindent
In view of these estimates, for the envisaged accuracy of $1\,\mu{\rm as}$ the quadrupole light deflection 
(\ref{source_5}) for sources in the solar system can be approximated by
\begin{eqnarray}
\delta \ve{k}_{\rm Q} &=&{1+\gamma\over 2}\, \frac{G}{c^2} \, \sum\limits_{A} \,
\ve{\alpha}_A^{\prime \prime}\,\frac{{\cal A}_A}{c} \,, 
\label{source_180}
\end{eqnarray}

\noindent
with ${\cal A}_A$ given by (\ref{source_10}) and $\ve{\alpha}_A^{\prime \prime}$ given by (\ref{source_30}).


\section{An upper estimate of the quadrupole light deflection for solar system objects\label{CriteriaSolar}}

The simplified expression of quadrupole light deflection for solar system objects (\ref{source_180}) it still 
complicated. In order to avoid evaluation of this term for each object in the data reduction, a simple
criterion is needed which allows one, with a few additional arithmetical operations, to judge if the quadrupole 
light deflection should be computed for a given source and for a given accuracy. To deduce such a criterion, 
we first consider the absolute value of the vectorial coefficient (\ref{source_70}) given by
\begin{eqnarray}
\left|\ve{\alpha}_A^{\prime \prime}\right| &=& M_A \, \left| J_2^A \right| \, P_A^2
\left(1 - \left(\ve{\sigma} \cdot \ve{e}_3\right)^2 \right) \,,
\label{criterion_source_5}
\end{eqnarray}

\noindent
so that an estimate of the absolute value of one term in (\ref{source_180}) is
\begin{eqnarray}
|\,\delta \ve{k}_{\rm Q}^A\,| &=& \frac{G}{c^2} \, M_A \, \left| J_2^A \right| \, P_A^2 \,
\left(1 - \left(\ve{\sigma} \cdot \ve{e}_3\right)^2 \right) \frac{|{\cal A}_A|}{c} \,,
\label{criterion_source_10}
\end{eqnarray}

\noindent
where the scalar coefficient $A_A$ is given in (\ref{source_10}). According to Eq.~(\ref{impact_k}) we have 
\begin{eqnarray}
\left(\ve{k} \times \ve{r}_{\source}^A\right)^2 &=& \left(\ve{k} \times \ve{r}_{\obs}^A\right)^2 
= d_A^2 + {\cal O} \left(c^{-2}\right)\,,
\label{criterion_source_15}
\end{eqnarray}

\noindent
and we obtain
\begin{eqnarray}
\frac{{\cal A}_A}{c} &=& \frac{1}{d_A^3} \, \frac{1}{R}
\left(\frac{1}{r_{\source}^A} \left( r_{\source}^A + \ve{k} \cdot \ve{r}_{\source}^A \right)^2
- \frac{1}{r_{\obs}^A} \left( r_{\obs}^A + \ve{k} \cdot \ve{r}_{\obs}^A \right)^2 \right)
\nonumber\\
\nonumber\\
&& + \frac{1}{\left(r_{\obs}^A\right)^3} \, \frac{1}{d_A^3}
\left( 2 \, r_{\obs}^A - \ve{k} \cdot \ve{r}_{\obs}^A \right)
\left( r_{\obs}^A + \ve{k} \cdot \ve{r}_{\obs}^A\right)^2 \,.
\label{criterion_source_20}
\end{eqnarray}

\noindent
Using $\ve{k} = \frac{\displaystyle \ve{R}}{\displaystyle R}$, and collecting all terms together we obtain
\begin{eqnarray}
\frac{{\cal A}_A (t_{\obs})}{c} &=& \frac{1}{d_A^3} \, \frac{1}{R^3} \,
(1\, -\,\cos \alpha)^2\,\left(2 \, \left(r_{\source}^A\right)^3 + \left(r_{\obs}^A\right)^2 \, r_{\source}^A 
+ 2 \, \left(r_{\source}^A\right)^2 \, r_{\obs}^A + \left(r_{\source}^A\right)^3 \, \cos \alpha \right).
\label{criterion_source_25}
\end{eqnarray}

\noindent
By means of the inequality (see Appendix \ref{appendix-Proof} for a proof)
\begin{eqnarray}
&& (1 - \cos \alpha)^2 \,
\frac{2 x^3 + x y^2 + 2 x^2 y + x^3 \cos \alpha}
{(x^2 + y^2 - 2  x  y  \cos \alpha)^{3/2}}
\le 3 \frac{x}{(x^2+y^2-2 x y \cos \alpha)^{1/2}} 
\frac{\sin^2 \alpha}{1 + \cos \alpha} \,,
\label{criterion_source_30}
\end{eqnarray}

\noindent
valid for any $x\ge0$ and $y\ge0$ and with the aid of
\begin{eqnarray}
\frac{r_{\source}^A}{R} \, \sin \alpha &=& \frac{d_A}{r_{\obs}^A} \,,
\label{criterion_source_32}
\end{eqnarray}

\noindent
we obtain the estimate,
\begin{eqnarray}
\frac{{\cal A}_A}{c} &\le& 3 \, \frac{1}{r_{\obs}^A} \,
\frac{1}{d_A^2} \, \frac{\sin\, \alpha}{1 + \cos \alpha}\,,
\label{criterion_source_35}
\end{eqnarray}

\noindent
and, therefore, with the aid of (\ref{source_180}) and (\ref{criterion_source_5}) we achieve
\begin{eqnarray}
| \delta {{\mbox{\boldmath $k$}}}_{\rm Q}^A | &\le&
{3\,(1+\gamma)\over 2}\,\frac{G M_A}{c^2}\,
\frac{1}{d_A^2} \, \frac{1}{r_{\obs}^A} \, \left| J_2^A \right| \, P_A^2 \,
\left(1 - \left(\ve{\sigma} \cdot \ve{e}_3\right)^2 \right) 
\frac{\sin\, \alpha}{1 + \cos \alpha}\,.
\label{criterion_source_40}
\end{eqnarray}

\noindent
This result can be related to the spherically symmetric part, given in (\ref{criterion_source_50}). By 
comparison between (\ref{criterion_source_40}) and (\ref{criterion_source_50}) we obtain the criterion
\begin{eqnarray}
\left|\,\delta \ve{k}_{\rm Q}^A \,\right| &\le&
\frac{3}{2} \, \frac{P_A^2}{d_A^2} \, \left| J_2^A \right|\,
\left( 1 - \left( \ve{\sigma} \cdot \ve{e}_3\right)^2 \right)\,\left|\,\delta \ve{k}_{\rm pN}^A\right| \,.
\label{criterion_source_55}
\end{eqnarray}

\noindent
Due to $1 \ge \left( \ve{\sigma} \cdot \ve{e}_3\right)^2 $,
the estimate (\ref{criterion_source_55}) can be further approximated by
\begin{eqnarray}
|\,\delta \ve{k}_{\rm Q}^A \, | &\le& \frac{3}{2} \, \left| J_2^A \right|\,
\frac{P_A^2}{d_A^2} \,|\,\delta \ve{k}_{\rm pN}^A | \,.
\label{criterion_source_60}
\end{eqnarray}

\noindent
This criterion relates the quadrupole light deflection of sources in the
solar system to the simpler case of spherically symmetric part.
For Gaia it is recommended to use (\ref{criterion_source_60}) as a criterion
if the quadrupole light deflection has to be calculated for a given
solar system object. The estimate of the monopole light deflection for solar system objects 
can be written as follows: 
\begin{eqnarray}
\left| \delta \ve{k}_{\rm pN}^A \right| &\le& 
{2\,(1+\gamma)\,GM_A\over c^2\,d_A}\le {2\,(1+\gamma)\,GM_A\over c^2\,P_A}\,,
\label{criterion_source_61}
\end{eqnarray}

\noindent
which can be used in the case when $|\,\delta \ve{k}_{\rm pN}^A\,|$ is not available; 
the proof of (\ref{criterion_source_61}) is straightforward by means of (\ref{criterion_source_50}).
From (\ref{criterion_source_10}) and (\ref{criterion_source_25}) one can directly see that
(for a proof see Appendix (\ref{appendix-Proof2}))
\begin{eqnarray}
\left| \delta \ve{k}_{\rm Q}^A \right| &\le&
2\,(1+\gamma)\,\frac{G M_A}{c^2}\,
\frac{P_A^2}{d_A^3} \, \left| J_2 \right| \le 2\,(1+\gamma)\,\frac{G M_A}{c^2\,P_A}\,\left| J_2 \right|\,,
\label{criterion_source_62}
\end{eqnarray}

\noindent
where we have used $P_A \le d_A$.


\section{Shapiro effect for solar system objects\label{Shapiro}}

In this Section, for reasons of completeness, the known analytical expressions of quadrupole Shapiro effect 
will be reconsidered, which is also of practical interest for astrometric missions in nearest future; 
e.g. BepiColumbo or Juno require modelling of the light travel time at the level of millimeters. 
From Eq.~(\ref{light_path_5}) we obtain for the Shapiro effect in post-Newtonian order the expression
\begin{eqnarray}
c \tau &=& R + c \sum\limits_i \delta \tau_i + {\cal O} \left(c^{-4}\right)\,,
\label{shapiro_10}
\end{eqnarray}

\noindent
where $c \tau = c (t_{\obs} - t_{\source})$ and 
\begin{eqnarray}
c\,\delta \tau_i &=& - \ve{k} \cdot \Delta \ve{x}_i (t_{\obs})\,.
\label{shapiro_20}
\end{eqnarray}

\noindent
In order to show (\ref{shapiro_10}) and (\ref{shapiro_20}) we have used (cf. Eq.~(23) of \cite{Klioner1})
\begin{eqnarray}
\ve{\sigma} &=& 
\ve{k} - R^{-1} \,\ve{k} \times \left(\sum\limits_i \Delta \ve{x}_i (t_{\obs})\times\ve{k}\right) 
+ {\cal O} \left(c^{-4}\right)\,.
\label{shapiro_15}
\end{eqnarray}
\noindent
The sum in (\ref{shapiro_10}) runs over the terms of the metric caused by the massive body 
(e.g. spherical symmetric term, quadrupole term and higher multipole terms etc.).
Here, for our purposes it will be sufficient to consider the spherical symmetric and quadrupole part. 
The spherical symmetric term of Shapiro effect is given by 
\begin{eqnarray}
c\,\delta \tau_{\rm pN} &=& - \ve{k} \cdot \Delta_{\rm pN} \,\ve{x} (t_{\obs}) 
= \sum \limits_A c\,\delta \tau_{\rm pN}^A \,,
\nonumber\\
\nonumber\\
c\,\delta \tau_{\rm pN}^A &=& \left(1+\gamma\right) \frac{G}{c^2}\,M_A \,
\log \frac{r_{\source}^A + r_{\obs}^A + R}{r_{\source}^A + r_{\obs}^A - R}\,,
\label{shapiro_27}
\end{eqnarray}

\noindent
where the sum runs over all massive bodies $A$ under consideration. 
The quadrupole term of Shapiro effect is given by 
\begin{eqnarray}
c\,\delta \tau_{\rm Q} &=& - \ve{k} \cdot \Delta_{\rm Q} \,\ve{x} (t_{\obs}) 
= \sum \limits_A c\,\delta \tau_{\rm Q}^A\,.
\label{shapiro_30}
\end{eqnarray}

\noindent
The expression $\Delta_{\rm Q}\,\ve{x}(t_{\obs})$ has been given in \cite{Klioner2}.
Accordingly, the Shapiro effect for quadrupole gravitational fields of one massive solar system body $A$ 
is given  by 
\begin{eqnarray}
c\,\delta \tau_{\rm Q}^A &=& \frac{1 + \gamma}{2} \,\frac{G}{c^2}\,
\left(\delta_A \, {\cal V}_A + \beta_A \, {\cal E}_A + \gamma_A \, {\cal F}_A \right),
\label{shapiro_50}
\end{eqnarray}

\noindent
with the scalar functions 
\begin{eqnarray}
{\cal E}_A &=& \frac{\ve{k} \cdot \ve{r}_{\source}^A}{\left(r_{\source}^A\right)^3}
- \frac{\ve{k} \cdot \ve{r}_{\obs}^A}{\left(r_{\obs}^A\right)^3}\,,
\label{shapiro_55}
\\
{\cal F}_A &=& d_A \left( \frac{1}{\left(r_{\source}^A\right)^3} - \frac{1}{\left(r_{\obs}^A\right)^3} \right)\,,
\label{shapiro_60}
\\
{\cal V}_A &=& - \frac{1}{d_A^2}
\left( \frac{\ve{k} \cdot \ve{r}_{\source}^A}{r_{\source}^A}-\frac{\ve{k}\cdot\ve{r}_{\obs}^A}{r_{\obs}^A}\right)\,,
\label{shapiro_65}
\end{eqnarray}

\noindent
and the time-independent scalar coefficients are 
\begin{eqnarray}
\beta_A &=& M_{i j}^A\,k_i\,k_j - M_{i j}^A\,\frac{d_A^i}{d_A}\,\frac{d_A^j}{d_A} \,,
\label{shapiro_70}
\\
\gamma_A &=& 2\,M_{i j}^A\,k^i\,\frac{d_A^j}{d_A} \,,
\label{shapiro_75}
\\
\delta_A &=& M_{i j}^A \,k_i\,k_j + 2\,M_{i j}^A\,\frac{d_A^i}{d_A}\,\frac{d_A^j}{d_A}\,.
\label{shapiro_80}
\end{eqnarray}

\noindent
Note, the impact vector can be computet by means of Eq.~(\ref{impact_k}).
In Appendix \ref{Proof_Shapiro_Effect} we show the following estimates:
\begin{eqnarray}
&& \frac{G}{c^2} \left|\,\delta_A\,{\cal V}_A\,\right|\le\frac{G\,M_A}{c^2}\,\left|J_2^A\right|\,
\frac{P_A^2}{d_A^2}\,,
\label{shapiro_140}
\\
&& \frac{G}{c^2} \left|\,\beta_A\,{\cal E}_A + \gamma_A\,{\cal F}_A\,\right| 
\le \frac{G\,M_A}{c^2}\,\left|J_2^A\right|\,
\left(\frac{P_A^2}{\left(r_{\source}^A\right)^2} + \frac{P_A^2}{\left(r_{\obs}^A\right)^2}\right)\,.
\label{shapiro_141}
\end{eqnarray}

\noindent
These estimates imply that (\ref{shapiro_50}) cannot be simplified for the general case.
Furthermore, from these relations, because of $P_A \le d_A, r_{\source}^A, r_{\obs}^A$, we conclude the inequality
\begin{eqnarray}
\left|\,c\,\delta \tau_{\rm Q}^A\,\right|
&\le& 3\,\left|J_2^A\right|\,\frac{G\,M_A}{c^2}\,,
\label{shapiro_150}
\end{eqnarray}

\noindent
which represents a strict upper bound of quadrupole Shapiro effect and 
improves the estimate given in Eq.~(47) in \cite{Klioner1}. This estimate implies that for quadrupole light
deflection there is a maximal numerical value which depends only on physical parameters of the massive body, 
but not on distance $d_A$. Numerical values of the estimate (\ref{shapiro_150}) for the giant planets are given 
in Table~\ref{table3}.

\begin{table}[t!]
\begin{tabular}{c | c | c | c | c | c}
&&&&&\\[-10pt]
Parameter &\hbox to 2mm{} Sun \hbox to 2mm{} &\hbox to 2mm{} Jupiter \hbox to 2mm{} & \hbox to 2mm{} Saturn \hbox to 2mm{}& \hbox to 2mm{} Uranus \hbox to 2mm{} & \hbox to 2mm{} Neptune \hbox to 2mm{} \\[3pt]
\hline
&&&&&\\[-10pt]
$3\,\left|J_2^A\right|\,\frac{\displaystyle G\,M_A}{\displaystyle c^2}$\ [${\rm mm}$]
& $0.89$ & $62.16$ & $20.68$ & $0.68$ & $0.81$\\[3pt]
\end{tabular}
\caption{Numerical values of estimate (\ref{shapiro_150}). For the Sun a value 
$J_2^{\odot} = 2\,\times 10^{-7}$ has been adopted \cite{J2_Sun}; $G M_{\odot}/c^2 = 1476\,{\rm m}$.}
\label{table3}
\end{table}


\section{Numerical tests\label{NumericalTests}}

The obtained simplified formulas given by Eq.~(\ref{light_185}) for
stars/quasars and by Eq.~(\ref{source_180}) for solar system objects and the
a priori criteria given for these objects by Eq.~(\ref{criterion_30}) and
(\ref{criterion_source_60}), respectively, have been incorporated into
the current reference C implementation of GREM as described by
\citet{Klioner:Blankenburg:2003} and \citet{Klioner:2003}.

Numerical experiments with the C implementation have confirmed the
correctness and the efficiency of the estimates and criteria. In the
experiments we used about $10^8$ objects (both randomly distributed
over the sky and specially generated to give grazing rays to the giant
planets). The results can be summarized as follows:

\begin{itemize}

\item[I.] Stars and quasars.

\begin{enumerate}

\item[--] The maximal difference between the full quadrupole deflection
formula (\ref{light_35}) and the simplified one (\ref{light_185})
amounts to $1.1\times10^{-10}$ \muas\ in good agreements with
(\ref{light_180}). Hence, the actual values of the neglected terms 
are in case of Gaia about 10--15 times less than given by
(\ref{light_180}).

\item[--] The upper estimate (\ref{criterion_30}) holds and is
attainable for randomly distributed sources.

\item[--] The mean value of the ratio between the actual value of the
quadrupole light deflection and its upper estimate (\ref{criterion_30})
amounts to 0.48 for randomly distributed sources that indicates the
high numerical efficiency of the estimate.

\end{enumerate}

\item[II.] Solar system objects.

\begin{enumerate}

\item[--] The maximal difference between the full quadrupole deflection
formula (\ref{source_5}) and the simplified one (\ref{source_180})
amounts to $0.0017$ \muas\ in good agreements with (\ref{source_175}).
Therefore, the actual values of the neglected terms are again about
10--15 times less than given by (\ref{source_175}).

\item[--] The upper estimate (\ref{criterion_source_60}) holds and is
attainable for randomly distributed sources.

\item[--] The mean value of the ratio between the actual value of the
quadrupole light deflection and its upper estimate (\ref{criterion_source_60})
amounts to 0.40 for randomly distributed sources that again indicates
the high numerical efficiency of the estimate.

\end{enumerate}

\item[III.] Quadrupole Shapiro effect.

\begin{enumerate}

\item[--] A strict upper estimate of quadrupole Shapiro effect is given in Eq.~(\ref{shapiro_150}).

\end{enumerate}

\end{itemize}

Implementation of the criteria (\ref{criterion_30}) and
(\ref{criterion_source_60}) has allowed to significantly reduce the
number of ``false alarms'' (cases for which the full quadrupole
deflection has been computed and turned out to be much smaller than the
requested goal accuracy). The ``false alarms'' were caused by the use
of a more primitive ad hoc criteria implemented in the C code of GREM
previously. This, in turn, slightly increases the performance of the
GREM implementation.


\section{Summary\label{Summary}}

Let us summarize the results of this report.

\begin{enumerate}
\item[1.] Quadrupole light deflection (\ref{light_35}) for stars
and quasars is approximated by (\ref{light_185}).
\item[2.] Eqs.~(\ref{criterion_30}) and (\ref{criterion_32}) can be used as an a priori criterion
if the quadrupole light deflection (\ref{light_185}) has to be computed for a given source.
\item[3.] Quadrupole light deflection (\ref{source_5}) for solar
system sources is approximated by (\ref{source_180}).
\item[4.] Eqs.~(\ref{criterion_source_60}) and (\ref{criterion_source_62}) can be used as an a priori
criterion if the quadrupole light deflection (\ref{source_180}) has to be computed for a given solar system object.
\item[5.] A strict criterion of quadrupole Shapiro effect has been given in Eq.~(\ref{shapiro_150}), which improves 
the estimate given in Eq.~(47) of \cite{Klioner1}.
\end{enumerate}

Our investigations, i.e. the simplified quadrupole formulas and the criteria, provide a highly time-efficient tool 
for data reduction on microarcsecond level of accuracy for Gaia mission.

\newpage

\appendix

\section{Proof of Eq.~(\ref{criterion_source_30})}\label{appendix-Proof}

In this Appendix we prove the inequality (\ref{criterion_source_30}). The latter can be rewritten as
\begin{eqnarray}
(1 - \cos \alpha)^2 \,
(2 x^3 + x\, y^2 + 2 x^2 \, y + x^3 \, \cos \alpha) &\le&
 3 \, x \, (x^2+y^2-2 x y\,\cos \alpha) \, (1 - \cos \alpha) \,.
\nonumber\\
\label{proof_5}
\end{eqnarray}

\noindent
Denoting $z=x/y$ we obtain the relation
\begin{eqnarray}
f_1 &\equiv&
- 1 - 2 \, z^2 + 2 \, z - \cos \alpha + 4 \, z \, \cos \alpha
- \cos^2 \alpha - z^2 \, \cos \alpha \, \le \, 0 \,.
\label{proof_10}
\end{eqnarray}

\noindent
To prove the inequality (\ref{proof_10}) it is sufficient to investigate the values of $f$ at extrema and at
the boundaries given by $z\ge0$ and $0\le\alpha\le\pi$. To determine the extrema of this function we set the 
first derivatives zero,
\begin{eqnarray}
f_{1\,,\,z} &=& -4 \, z + 2 + 4 \, \cos \alpha - 2 \, z \, \cos \alpha
\, = \, 0 \,,
\label{proof_15}
\\
f_{1\,,\,{\alpha}} &=& \sin\, \alpha \,
(1 - 4 \, z + 2 \, \cos \alpha + z^2) \, = \, 0 \,.
\label{proof_20}
\end{eqnarray}

\noindent
The only solutions of the coupled system (\ref{proof_15}), (\ref{proof_20}) are
\begin{eqnarray}
P_1 &=& (z = 0, \alpha = \frac{2}{3} \, \pi) \,,
\quad P_2 \, = \, (z = 1, \alpha = 0) \,.
\label{proof_30}
\end{eqnarray}

\noindent
The values of $f_1$ at these points are
\begin{eqnarray}
f_1 (P_1) &=&  - \frac{3}{4} \,, \quad f_1 (P_2) \, = \, 0 \,.
\label{proof_40}
\end{eqnarray}

\noindent
At the boundaries we get
\begin{eqnarray}
f_1 (z=0) &=&  -1 - \cos \alpha - \cos^2 \alpha \, < 0 \,,
\label{proof_45}
\\
f_1 (z \rightarrow \infty) &=&  - z^2 (2 + \cos \alpha) \, < 0 \,,
\label{proof_50}
\\
f_1 (\alpha = 0) &=&  - 3 (1 - z)^2 \, \le 0 \,,
\label{proof_55}
\\
f_1 (\alpha = \pi) &=& -z (z + 2) \, \le \, 0 \,.
\label{proof_60}
\end{eqnarray}

\noindent
From the results (\ref{proof_30}) - (\ref{proof_60}) we conclude the validity
of the inequality (\ref{proof_10}) and (\ref{proof_5}).


\section{Proof of Eq.~(\ref{criterion_source_62})}\label{appendix-Proof2}

From (\ref{criterion_source_10}) it is clear that (\ref{criterion_source_62}) is true if for ${\cal A}_A$ defined 
by (\ref{criterion_source_25}) one has
\begin{equation}
\label{A_A-estimate}
\frac{{\cal A}_A (t_{\obs})}{c} \le \frac{4}{d_A^3} \,.
\end{equation}

\noindent
To prove this it is sufficient to demonstrate that
\begin{eqnarray}
\label{A_A-estimate-rest}
\frac{1}{R^3} \,
(1 \, - \, \cos \alpha)^2 \,
\left(2 \, \left(r_{\source}^A\right)^3 + \left(r_{\obs}^A\right)^2 \,r_{\source}^A 
+ 2 \, \left(r_{\source}^A\right)^2 \, r_{\obs}^A + \left(r_{\source}^A\right)^3 \,
\cos \alpha \right)&\le&4 \,,
\end{eqnarray}

\noindent
or introducing again $z=r_{\source}^A/r_{\obs}^A$
\begin{eqnarray}
\label{A_A-estimate-rest-z}
f_2\equiv{(1 - \cos\alpha)^2 \, z \,
\left((1+z)^2+z^2\,(1+\cos\alpha) \right)
\over 4\,{\left(1+z^2-2z\cos\alpha\right)}^{3/2}}
&\le&1\,.
\end{eqnarray}

\noindent
The derivatives of $f_2$ with respect to $\alpha$ and $z$ vanish simultaneously only for $\alpha=0$ which 
is one of the boundaries. At the boundaries we get
\begin{eqnarray}
f_2 (z=0) &=&  0 \,,
\label{proof_145}
\\
\lim_{z \rightarrow \infty}\,f_2 &=&  \frac{1}{4}\,(1-\cos\alpha)^2\,(2+\cos\alpha)\le1 \,,
\label{proof_150}
\\
f_2 (\alpha = 0) &=&0 \,,
\label{proof_155}
\\
f_2 (\alpha = \pi) &=& {z\over{1+z}}\le1 \,.
\label{proof_160}
\end{eqnarray}

\noindent
From this we conclude that (\ref{A_A-estimate-rest}) and, therefore,
(\ref{A_A-estimate}) and (\ref{criterion_source_62}) are valid.


\section{Proof of inequalities (\ref{shapiro_140}) and (\ref{shapiro_141})\label{Proof_Shapiro_Effect}}

First, we consider the inequality (\ref{shapiro_140}), which is given by 
\begin{eqnarray}
\frac{G}{c^2}\,\left|\,\delta_A\,{\cal V}_A\,\right| &=& 
\frac{1}{d_A^2}\,\left|\,
\left( M_{i j}^A \,k_i\,k_j + 2\,M_{i j}^A\,\frac{d_A^i}{d_A}\,\frac{d^j}{d_A}\right) 
\left(\frac{\ve{k} \cdot \ve{r}_{\source}^A}{r_{\source}^A}-\frac{\ve{k}\cdot\ve{r}_{\obs}^A}{r_{\obs}^A}\right) 
\,\right|
\label{appendix_shapiro_1}
\\
\nonumber\\
&\le& \frac{1}{d_A^2}\,\left|\,M_{i j}^A \,k_i\,k_j + 2\,M_{i j}^A\,\frac{d_A^i}{d_A}\,\frac{d_A^j}{d_A}\,\right|\,,
\label{appendix_shapiro_2}
\end{eqnarray}

\noindent
where the inequality in (\ref{appendix_shapiro_2}) is obvious. 
By inserting (\ref{light_95}) into (\ref{appendix_shapiro_2}) we obtain 
\begin{eqnarray}
\frac{G}{c^2}\,\left|\,\delta_A\,{\cal V}_A\,\right| &\le& 
\frac{G\,M_A}{c^2}\,\left|J_2^A\right|\,\frac{P_A^2}{d_A^2}\,
\left| \, 1 - \left(\ve{k}\cdot\ve{e}_3\right)^2 - 2\,\frac{\left(\ve{d}_A \cdot \ve{e}_3\right)^2}{d_A^2}\,\right|\,.
\label{appendix_shapiro_5}
\end{eqnarray}

\noindent
Using the inequality  
\begin{eqnarray}
\left| \, 1 - \left(\ve{k}\cdot\ve{e}_3\right)^2 - 2\,\frac{\left(\ve{d}_A \cdot \ve{e}_3\right)^2}{d_A^2}\,\right| 
&\le& 1\,,
\label{appendix_shapiro_10}
\end{eqnarray}

\noindent
which can be shown by introducing spherical coordinates and taking into account that in post-Newtonian order 
$\ve{k}$ and $\ve{d}_A$ are perpendicular to each other (cf. proof of Eq.~(\ref{appendix_shapiro_45})), we find 
\begin{eqnarray}
\frac{G}{c^2}\,\left|\,\delta_A\,{\cal V}_A\,\right| &\le& 
\frac{G\,M_A}{c^2}\,\left|\,J_2^A\,\right|\,\frac{P_A^2}{d_A^2}\,.
\label{appendix_shapiro_15}
\end{eqnarray}

\noindent
Now, we consider the inequality (\ref{shapiro_141}). From the definitions of the functions (\ref{shapiro_55}) and 
(\ref{shapiro_60}) and of the scalar coefficients (\ref{shapiro_70}) - (\ref{shapiro_75}), we obtain 
\begin{eqnarray}
&& \frac{G}{c^2}\,\left|\,\beta_A\,{\cal E}_A + \gamma_A\,{\cal F}_A\,\right| 
\nonumber\\
\nonumber\\
&=& \left|\,\left( M_{i j}^A\,k^i\,k^j - M_{i j}^A\,\frac{d_A^i}{d_A}\,\frac{d_A^j}{d_A}\right) 
\left(\frac{\ve{k}\cdot\ve{r}_{\source}^A}{\left(r_{\source}^A\right)^3} - 
\frac{\ve{k}\cdot\ve{r}_{\obs}^A}{\left(r_{\obs}^A\right)^3}\right)
+ 2\,M_{i j}^A\,k^i\,\frac{d_A^j}{d_A} 
\left(\frac{d_A}{\left(r_{\source}^A\right)^3} - \frac{d_A}{\left(r_{\obs}^A\right)^3}\right)\,\right| 
\nonumber\\
\nonumber\\
&\le& \frac{f_3}{\left(r_{\source}^A\right)^2} + \frac{f_4}{\left(r_{\obs}^A\right)^2}\,,
\label{appendix_shapiro_20}
\end{eqnarray}

\noindent
where the functions are given by 
\begin{eqnarray}
f_3 &=& \left|\,\left( M_{i j}^A\,k^i\,k^j - M_{i j}^A\,\frac{d_A^i}{d_A}\,\frac{d_A^j}{d_A} \right)
\frac{\ve{k}\cdot\ve{r}_{\source}^A}{r_{\source}^A} 
+ 2\,M_{i j}^A\,k^i\,\frac{d_A^j}{d_A}\,\frac{d_A}{r_{\source}^A}\,\right|\,,
\label{appendix_shapiro_25}
\\
\nonumber\\
f_4 &=& \left|\,\left( M_{i j}^A\,k^i\,k^j - M_{i j}^A\,\frac{d_A^i}{d_A}\,\frac{d_A^j}{d_A} \right)
\frac{\ve{k}\cdot\ve{r}_{\obs}^A}{r_{\obs}^A} 
+ 2\,M_{i j}^A\,k^i\,\frac{d_A^j}{d_A}\,\frac{d_A}{r_{\obs}^A}\,\right|\,.
\label{appendix_shapiro_30}
\end{eqnarray}

\noindent
Now we consider function $f_3$ (proof for function $f_4$ is very similar). 
Inserting (\ref{light_95}) into (\ref{appendix_shapiro_25}) yields 
\begin{eqnarray}
f_3 &=& \frac{G\,M_A}{c^2}\,P_A^2\,\left|J_2^A\right|\,
\bigg| \left(\frac{\left(\ve{d}_A \cdot \ve{e}_3\right)^2}{d_A^2} - \left(\ve{k} \cdot\ve{e}_3\right)^2\right) 
\frac{\ve{k} \cdot \ve{r}_{\source}^A}{r_{\source}^A} - 2\,\left(\ve{k}\cdot\ve{e}_3\right)
\frac{\left(\ve{d}_A \cdot \ve{e}_3\right)}{d_A}\,\frac{d_A}{r_{\source}^A}\bigg|\,.
\label{appendix_shapiro_35}
\end{eqnarray}

\noindent
Using the inequality (see below) 
\begin{eqnarray}
h &=& \bigg| \left(\frac{\left(\ve{d}_A \cdot \ve{e}_3\right)^2}{d_A^2} - \left(\ve{k} \cdot\ve{e}_3\right)^2\right)
\frac{\ve{k} \cdot \ve{r}_{\source}^A}{r_{\source}^A} - 2\,\left(\ve{k}\cdot\ve{e}_3\right) \frac{\left(\ve{d}_A 
\cdot \ve{e}_3\right)}{d_A}\,\frac{d_A}{r_{\source}^A}\bigg| \le 1\,,
\label{appendix_shapiro_45}
\end{eqnarray}

\noindent
we find 
\begin{eqnarray}
\frac{G}{c^2} \left|\,\beta_A\,{\cal E}_A + \gamma_A\,{\cal F}_A\,\right|
\le \frac{G\,M_A}{c^2}\,\left|J_2^A\right|\,
\left(\frac{P_A^2}{\left(r_{\source}^A\right)^2} + \frac{P_A^2}{\left(r_{\obs}^A\right)^2}\right)\,,
\label{appendix_shapiro_50}
\end{eqnarray}

\noindent
which is just relation (\ref{shapiro_141}). 

In order to complete the proof of (\ref{appendix_shapiro_50}), we still have to show relation 
(\ref{appendix_shapiro_45}). For that we introduce spherical coordinates as follows: 
$\ve{d}_A = \left(0,0,d_A\right)^{\rm T}$, $\ve{k} = \left(\cos \phi_1\,,\, \sin \phi_1\,,\, 0\right)^{\rm T}$ 
and $\ve{e}_3 = \left(\sin \theta \,\cos \phi_2\,,\, \sin \theta\, \sin \phi_2\,,\,\cos \theta\right)^{\rm T}$, 
where we have taken into account that in post-Newtonian order $\ve{d}_A$ and $\ve{k}$ are perpendicular 
to each other. Accordingly, we have $\ve{d}_A\cdot \ve{e}_3 = d_A\,\cos \theta$ and 
$\ve{k}\cdot\ve{e}_3 = \sin \theta \left(\cos \phi_1\,\cos \phi_2 + \sin \phi_1\,\sin \phi_2\right)$, 
and by inserting into Eq.~(\ref{appendix_shapiro_45}) we obtain
\begin{eqnarray}
h &=& \left|\left( \cos^2 \theta - \sin^2 \theta\,\cos^2 \alpha \right) \cos \Phi  
- 2 \left(\cos \theta\,\sin \theta\,\cos \alpha \right) \sin \Phi \right| \,,
\label{appendix_shapiro_55}
\end{eqnarray}

\noindent
where we have used $\frac{\displaystyle \ve{k}\cdot\ve{r}_{\source}^A}{\displaystyle r_{\source}^A} = \cos \Phi$ 
and $\frac{\displaystyle d_A}{\displaystyle r_{\source}^A} = \sin \Phi$, and the addition theorem of  
cosine and sine function: $\cos \phi_1\,\cos \phi_2 + \sin \phi_1\,\sin \phi_2 = \cos \alpha$ 
where $\alpha= \phi_1 - \phi_2$. The expression in (\ref{appendix_shapiro_55}) can be estimated by 
$\left|\,A\,\cos \Phi + B\,\sin \Phi\,\right| \le \sqrt{A^2 + B^2}$; accordingly we obtain:
\begin{eqnarray}
h &\le& \cos^2 \theta + \sin^2\theta \,\cos^2 \alpha \le \cos^2 \theta + \sin^2\theta = 1\,.
\label{appendix_shapiro_60}
\end{eqnarray}

\noindent
Thus, we have shown relation (\ref{appendix_shapiro_45}).
 
\end{document}